\documentclass[aps,pra,twocolumn]{revtex4}
\newcommand {\be}{\begin{equation}}
\newcommand {\ee}{\end{equation}}
\newcommand {\bey}{\begin{eqnarray}}
\newcommand {\eey}{\end{eqnarray}}

\usepackage{epsfig}
\usepackage{amsmath}
\usepackage{mathbbol}

\usepackage{hyperref}

\begin{document}

\title{Exponential communication gap between weak and strong classical
simulations of quantum communication}

\author{Alberto Montina}
\affiliation{Perimeter Institute for Theoretical Physics, 31 Caroline Street North, Waterloo, 
Ontario N2L 2Y5, Canada \\ and \\
Facolt\`a di Informatica, 
Universit\`a della Svizzera Italiana, Via G. Buffi 13, 6900, Lugano, Switzerland
}

\date{\today}

\begin{abstract}
The most trivial way to simulate classically the communication of a quantum state is to transmit 
the classical description of the quantum state itself. However, this requires an infinite amount of 
classical communication if the simulation is exact. A more intriguing and potentially less demanding 
strategy would encode the full information about the quantum state into the probability distribution 
of the communicated variables, so that this information is never sent in each single shot. This 
kind of simulation is called weak, as opposed to strong simulations, where the 
quantum state is communicated in individual shots. In this paper, we 
introduce a bounded-error weak protocol for simulating the communication of an arbitrary 
number of qubits and a subsequent two-outcome measurement consisting of an arbitrary pure state 
projector and its complement. This protocol requires an amount of classical communication 
independent 
of the number of qubits and proportional to $\Delta^{-1}$, where $\Delta$ is the error and a 
free parameter of the protocol. Conversely, a bounded-error strong protocol requires
an amount of classical communication growing exponentially with the number of qubits for a
fixed error. Our result improves a previous protocol, based on the Johnson-Lindenstrauss lemma, 
with communication cost scaling as $\Delta^{-2}\log\Delta^{-1}$.
\end{abstract}

\maketitle

\section{Introduction}

In the framework of quantum theory, the quantum state of a system does not represent any physical
attribute of the system itself and just provides an operational (classical) description of a preparation 
procedure. This framework displays a dichotomy between the macroscopic realm, described through a 
classical language, and the quantum world, whose actual physical state is not provided by the 
formalism in terms of a classical picture. This difference of representation, which is the core 
of the measurement problem in quantum theory, has actually a simple solution. The quantum state,
possibly supplemented by additional variables, can be interpreted as a physical classical state 
conditioning the outcomes of measurements. Such a solution of the measurement problem is employed, 
for example, by collapse theories and Bohm mechanics. This ``cheap'' way to introduce a classical language 
in the description of a quantum system is also the most trivial way to simulate classically quantum 
communication. Indeed a process of quantum state preparation, its transmission through a quantum 
channel and subsequent measurement can be classically simulated by directly transmitting the classical 
description of the quantum state itself. Classical theories promoting the quantum state to the rank 
of a physical variable are called $\psi$-ontic (in Greek, {\it ontos} means {\it that which is}).

In the recent years, there has been an increasing interest for an alternative class of hypothetical 
classical theories, called $\psi$-epistemic~\cite{hardy,monti0,spekkens,monti1,monti2,monti3,
monti4,bartlett,monti5,pusey,lewis,colbeck,schloss,hardy1,montina}.
In their framework, the information about the quantum state 
is not stored in the classical state of each single quantum system, but it is encoded in the probability
distribution of the classical state (also called {\it ontic state}). 
In other words, in a $\psi$-epistemic theory, the quantum state represents statistical knowledge about the 
ontic state (in Greek, {\it episteme} means {\it knowledge}).

In quantum computer science, $\psi$-epistemic and $\psi$-ontic simulations are known as weak and strong
simulations~\cite{nest}, respectively. The task of a strong simulation of a quantum measurement is to 
evaluate the outcome probabilities of every measurement outcomes with a possible bounded error. 
In other words, the task is to evaluate the quantum state after some processing. Conversely, a single 
shot of a weak simulation just generates a measurement outcome according to its quantum probability. 
A weak simulation is more
similar to the actual experimental scenario that is simulated, where the final quantum state of
a single system cannot be directly measured, but it is tomographically reconstructed through many 
repetitions of the same experimental procedure.

As discussed in Ref.~\cite{montina}, $\psi$-epistemic theories have an important role in quantum 
communication and are related to a very practical question: how many classical bits of communication 
are required to simulate exactly the communication of qubits? 
If only $\psi$-ontic simulations were feasible, then the communication cost 
would be trivially infinite, as a $\psi$-ontic simulation is carried out by communicating the full 
infinite information about the quantum state. Thus, it is clear that a protocol that classically 
simulates quantum communication through a finite amount of classical information (called, more 
concisely, finite communication protocol or FC protocol) is a kind of $\psi$-epistemic (weak) 
protocol.
Furthermore, since the mutual information, say $I_m$, between the quantum state and the ontic
state is not greater than the communication cost, a FC protocol has $I_m$ finite. In 
Ref.~\cite{montina}, we called $\psi$-epistemic models with finite $I_m$ {\it completely 
$\psi$-epistemic}. Thus, a FC protocol is also completely $\psi$-epistemic. We proved that also the 
opposite is somehow true~\cite{montina}. More precisely, we showed that a completely 
$\psi$-epistemic protocol can be turned into a FC protocol. The communication cost, say $\cal C$, of 
the derived protocol is essentially given by the mutual information $I_m$ between the quantum
state and the ontic state of the parent protocol.
Indeed, using a recent result~\cite{harsha}, we showed that
$$
I_m\le{\cal C}\le I_m+2\log_2(I_m+1)+2\log_2e
$$
(the second inequality can be strengthened under a suitable condition, as pointed out later). 
Furthermore, there is a procedure that turns parallel $\psi$-epistemic simulations into a global 
protocol with asymptotic communication cost, say ${\cal C}^{asym}$, per simulation
exactly equal to the mutual information, as a consequence of the reversed 
Shannon theorem~\cite{rev_shannon}, in the form stated in Ref.~\cite{winter}.

Thus, the problem of finding a FC protocol is exactly equivalent to the problem of finding a completely
$\psi$-epistemic protocol, since a FC protocol is completely $\psi$-epistemic and, moreover, a 
completely $\psi$-epistemic protocol can be always turned into a FC protocol. 
This procedure of turning a broader class of protocols into a subclass simplifies 
the task of deriving FC protocols, as point out in Ref.~\cite{montina}.

Completely $\psi$-epistemic models are known only for single qubits. Consequently, only FC 
protocols simulating the communication of single qubits are known. Toner and Bacon proved that
the communication of two classical bits is sufficient to simulate the communication of a
qubit~\cite{toner}. In the case of many simulations performed in parallel, the asymptotic 
communication cost can be compressed to about $1.279$ bits~\cite{montina}. 
In this paper, we introduce a bounded-error $\psi$-epistemic protocol for simulating the
communication of an arbitrary number of qubits, followed by a measurement consisting of an 
arbitrary pure state projector, say $|\phi\rangle\langle\phi|$, and its complement, 
$\mathbb{1}-|\phi\rangle\langle\phi|$.
The error is a free parameter of the protocol and can be arbitrarily small. In the limit case 
of zero error, the model is $\psi$-ontic. Using the aforementioned procedure introduced 
in Ref.~\cite{montina},
we then derive an approximate FC protocol simulating the communication of $n$ qubits. 
The amount of required classical communication is independent of $n$ 
and inversely proportional to the worst-case error, say $\Delta$, in the limit $n\gg\log\Delta^{-1}$.
Conversely a bounded-error approximation of a brute force $\psi$-ontic simulation 
requires an amount
of classical communication growing exponentially with the number of qubits for a fixed $\Delta$.
Our protocol improves a previous protocol, based on the Johnson-Lindenstrauss lemma, 
with communication cost scaling as $\Delta^{-2}\log\Delta^{-1}$~\cite{brandao}.

The paper is organized as follows. In section~\ref{class_sim}, we introduce the general
structure of a classical simulation of a quantum channel. We then define the communication
cost of the classical simulation and the communication complexity of a quantum channel. 
These definitions, as well as the definition of completely $\psi$-epistemic protocols,
is slightly different from that given in Ref.~\cite{montina}. In the previous definition,
the communication cost was a function of the quantum state probability distribution,
which needed to be specified. The new definition does not have this dependence.
The definition of $\psi$-epistemic and $\psi$-ontic protocols is also generalized to 
the case of approximate simulations.
In section~\ref{procedure}, the procedure of Ref.~\cite{montina} is discussed with
slight changes reflecting the different definition of communication cost.
In Section~\ref{protocol}, we use this procedure to derive the approximate FC protocol. Finally
we compare the derived protocol with a bounded-error $\psi$-ontic protocol.

\section{Classical simulation of quantum channels}
\label{class_sim}
\begin{figure}
\epsfig{figure=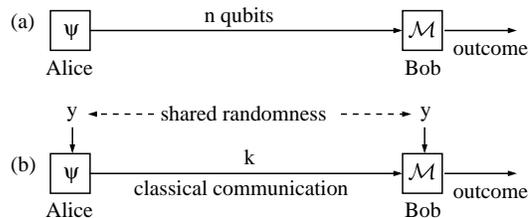,width=7.cm}
\caption{(a) Schematic representation of a two-party scenario with one-way quantum communication.  
(b) Quantum communication is replaced by classical communication and the two parties are allowed to 
use shared randomness, $y$.}
\label{fig1}
\end{figure}

A classical simulation of a quantum channel simulates more correctly a process of preparation,
transmission through the channel and measurement of a quantum state. We will consider only noiseless
quantum channels. The scenario that is classically simulated is illustrated in Fig.~\ref{fig1}a.
A party, say Alice, prepares some qubits in a quantum state 
$|\psi\rangle$ and sends them to another party, say Bob. Bob then generates an outcome by performing 
some measurement ${\cal M}\equiv\{\hat E_1,\hat E_2\dots\}$, where $\hat E_i$
are positive semidefinite self-adjoint operators labeling events of the measurement $\cal M$.
Notice that Alice has a classical description of the quantum state $|\psi\rangle$. In a more 
complicate scenario, which will not be discussed, Alice could perform some operations on qubits
received from a third party.

A classical simulation of the two-party quantum scenario is illustrated in Fig.~\ref{fig1}b. Instead 
of preparing the qubits in the state $|\psi\rangle$, Alice generates a classical variable $k$ with a 
probability $\rho(k|y,\psi)$ depending on the quantum state and a possible random variable, $y$, 
shared with Bob. 
Thus, there is a mapping from the quantum state to a probability distribution of $k$,
\be\label{map_psi_k}
|\psi\rangle\xrightarrow{y} \rho(k|y,\psi).
\ee
The variable $y$ is generated according to a probability distribution $\rho(y)$. 
The value of $k$ is communicated by Alice to Bob. Finally, Bob generates an outcome $\hat E_i$ with 
a probability $P(\hat E_i|k,y,{\cal M})$. The protocol simulates exactly the quantum channel if 
the probability of $\hat E_i$ given $|\psi\rangle$ is equal to the quantum probability,
that is, if
\be
\int dk\int dy P(\hat E_i|k,y,{\cal M}) \rho(k|y,\psi)\rho(y)=\langle\psi|\hat E_i|\psi\rangle.
\ee

If the variable $k$ takes a uncountably infinite number of values, the communication cost is
infinite. Conversely, if $k$ is a discrete variable, the communication cost, say $\cal C$,
is defined as the maximum, over the space of probability distributions $\rho(\psi)$, of
the Shannon entropy of the distribution $\rho(k|y)$ averaged over $y$, that is,
\be
{\cal C}\equiv\max_{\rho(\psi)}\left\{\int dy\rho(y)\left[-\sum_k \rho(k|y)\log_2\rho(k|y)\right]\right\}.
\ee
This definition differs from that given in Ref.~\cite{montina}, where the cost was not maximized
and was a function of the distribution $\rho(\psi)$, which needed to be specified.
We define the {\it communication complexity} (denoted by ${\cal C}_{min}$) of a quantum channel as the 
minimal amount of classical communication $\cal C$ required by an exact classical simulation of the
quantum channel. 

Shannon's source coding theorem~\cite{cover} establishes an operational meaning of $\cal C$. Suppose 
that $M$ independent simulations of $M$ quantum channels are performed in parallel. Let $k_i$ be the 
variable prepared with probability $\rho(k_i|y,\psi_i)$, where $|\psi_i\rangle$ is the quantum
state prepared for the $i$-th quantum channel. Instead of communicating directly the variables $k_i$,
we can encode them into a global $k$, so that
the average number of communicated bits per simulation approaches $\cal C$ as $M$ goes to infinity.
If the compression code is independent of the probability distribution $\rho(\psi)$, the compression
rate is minimal, as stated by Shannon's theorem.
In this parallelized protocol, the variables $k_i$ are first generated 
according to the one-shot protocol and then compressed into the global variable $k$. However, it is 
possible to envisage a larger set of communication protocols where the global variable $k$ is directly 
generated from the quantum states prepared in each single simulation. 
In other words, the probability distribution $\rho(k|y,\psi)$ of a single simulation is replaced by 
a probability distribution, say $\rho(k|y,\psi_1,\psi_2,\dots,\psi_{M})$, depending on the whole
set of $M$ prepared quantum states. Thus, we have the mapping 
\be
\{|\psi_1\rangle,\dots,|\psi_M\rangle\}\xrightarrow{y}\rho(k|y,\psi_1,\psi_2,\dots,\psi_{M}),
\ee
which replaces the single-shot mapping~(\ref{map_psi_k}).
The asymptotic communication cost, ${\cal C}^{asym}$, is the cost of this 
parallelized simulation divided by $M$, for $M$ going to infinity. 
We define the {\it asymptotic communication complexity}, ${\cal C}_{min}^{asym}$, of a quantum 
channel as the minimal asymptotic communication cost required for simulating the channel.
Since the set of protocols working for parallel simulations is larger than the set of protocols
obtained by just compressing the communication of independent one-shot protocols, it is clear that
\be
{\cal C}_{min}^{asym}\le {\cal C}_{min}.
\ee

A classical channel $x_1\rightarrow x_2$ from a stochastic variable $x_1$ to $x_2$ is defined
by a conditional probability $\rho(x_1|x_2)$. The capacity of the channel is the maximum of
the mutual information between $x_1$ and $x_2$ over the space of the input probability 
distributions $\rho(x_1)$~\cite{cover}. The mutual information of two variables $x_1$ and
$x_2$ with probability distribution $\rho(x_1,x_2)$ is 
$$
I(X_1;X_2)=\sum_{x_1,x_2}\rho(x_1,x_2)\log_2\frac{\rho(x_1,x_2)}{\rho(x_1)\rho(x_2)}.
$$
Here, the capital letters refer to the stochastic variables, whereas their lower case refers to 
value taken by the variables. Whenever there is no ambiguity, we just use the lower case also for 
the stochastic variables themselves. 
From the chain rule
\be
I(K,Y;\Psi)=I(Y;\Psi)+I(K;\Psi|Y)
\ee
and the fact that $|\psi\rangle$ and $y$ are uncorrelated, we have that
\be\label{from_chain}
I(K,Y;\Psi)=I(K;\Psi|Y).
\ee
The mutual information $I(K;\Psi|Y)$, for any $\rho(\psi)$, is smaller than or equal to the 
communication cost $\cal C$. Thus, from Eq.~(\ref{from_chain}), we have that
\be\label{entro_mutual}
{\cal C}\ge C(K,Y|\Psi),
\ee
where $C(K,Y|\Psi)$ is the capacity of the channel $|\psi\rangle\rightarrow\{k,y\}$ from the
quantum state to the classical variables of the protocol.

As said in the introduction, an exact $\psi$-ontic protocol trivially simulates a quantum channel by 
sending 
the full infinite information about the quantum state. Conversely, in a $\psi$-epistemic theory, this 
information is encoded in the statistical distribution of the communicated variable. Thus, it is clear
that an exact FC protocol is also a $\psi$-epistemic protocol. In the class of $\psi$-epistemic
protocols, there is a interesting subclass of protocols that we call {\it completely
$\psi$-epistemic} and characterized by the additional property that the channel capacity
$C(K;Y|\Psi)$ is finite. This condition is slightly stronger than the finiteness of the mutual 
information $I(K;Y|\Psi)$ for some fixed $\rho(\psi)$, used in Ref.~\cite{montina}.

\begin{figure}
\epsfig{figure=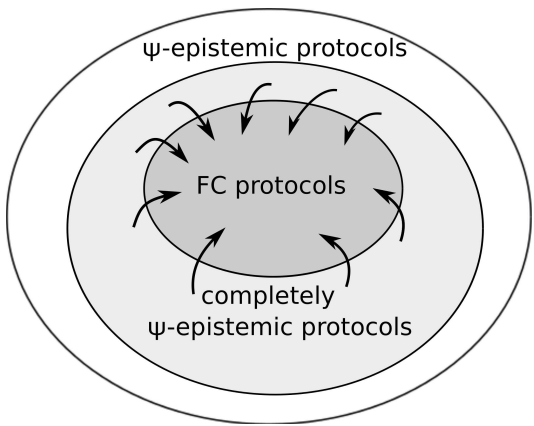,width=7.cm}
\caption{In $\psi$-epistemic protocols, the full information about the quantum state is not communicated 
in each single shot, but it is encoded in statistical distribution of the communicated variable. Completely 
$\psi$-epistemic protocols are characterized by the additional property that the capacity of the channel
$|\psi\rangle\rightarrow\{k,y\}$ is finite. There is a procedure that turns any completely
$\psi$-epistemic protocol into a FC protocol. This procedure is represented by the arrows.}
\label{fig2}
\end{figure}

It is clear from Eq.~(\ref{entro_mutual}) that a FC protocol is also a completely 
$\psi$-epistemic protocol. In the next section, we will show that also the opposite is somehow
true. More precisely, we will show that there is procedure turning a completely $\psi$-epistemic
protocol into a FC protocol. The hierarchy of the aforementioned classes is schematically
represented in Fig.~\ref{fig2}.

This hierarchy is broken in the case of bounded-error protocols. Indeed, a bounded-error
$\psi$-ontic protocol can be also a FC protocol. Thus, a bounded-error FC protocol in
not necessarily $\psi$-epistemic. We define an approximate $\psi$-ontic protocol as
follows.
First, Alice approximates the quantum state $|\psi\rangle$ with an element, say 
$|\psi_{sub}\rangle$, of a subset of vectors. It is important to stress that the mapping
\be
|\psi\rangle\rightarrow|\psi_{sub}\rangle
\ee
does not depend on any stochastic variable, that is, there a unique $|\psi_{sub}\rangle$
for each $|\psi\rangle$.
Then, like in an exact $\psi$-ontic protocol, the full information about the $|\psi_{sub}\rangle$
is sent to the receiver. 
In this protocol, there is not an encoding of the quantum state information in the probability 
distribution of the communicated variable, this information is just partially erased to make it
finite. By definition, an approximate $\psi$-epistemic protocol is 
any protocol that is not $\psi$-ontic. A bounded-error protocol generates the outcomes of
the simulated measurement in accordance to the quantum probabilities with an error that is 
bounded by an arbitrarily small constant, which is a free parameter of the protocol. 
In the case of $\psi$-ontic simulations, this
means that Bob has to receive, in a single shot, sufficient information so that he
is able to evaluate the probability of any event with a bounded error.

\section{FC protocols from completely $\psi$-epistemic protocols}
\label{procedure}

We now describe the procedure introduced in Ref.~\cite{montina} for generating a FC 
protocol from a completely $\psi$-epistemic protocol. This procedure is a consequence of the
reverse Shannon theorem~\cite{rev_shannon} and its one-shot version~\cite{harsha}. 
Given $M$ copies of a channel $x\rightarrow y$, defined by the conditional probability 
$\rho(y|x)$ and with capacity $C_{ch}$, the reverse Shannon theorem states that they
can be replaced by a global noiseless channel whose capacity is equal to $M C_{ch}+o(M)$,
provided that the sender and receiver share some random variable.
In other words the asymptotic communication cost of a parallel simulation of many copies of a 
channel $x\rightarrow y$ is equal to $C_{ch}$
A one-shot version of this theorem was recently reported in Ref.~\cite{harsha}.
For independently simulated realizations, we have that
\be\label{cost_one-shot}
C_{ch}\le{\cal C}\le C_{ch}+2\log_2(C_{ch}+1)+2\log_2e.
\ee
Thus, the communication cost is $C_{ch}$ plus a possible small additional 
cost that does not grow more than the logarithm of $C_{ch}+1$. The second inequality was
proved using an improved version of the rejection sampling method. If the probability
distribution $\rho(y|x)$ and the distribution $\rho(y)=\sum_x \rho(y|x)\rho(x)$, 
obtained by maximizing the mutual information, are uniform in their support, it is
possible to prove, using the standard rejection sampling method~\cite{rsm}, the 
stronger constraint
\be\label{cost_uniform}
C_{ch}\le{\cal C}\le C_{ch}+\log_2e.
\ee

These results have an immediate application to the problem of deriving FC protocols from 
completely $\psi$-epistemic protocols. In general, a $\psi$-epistemic protocol can have an 
infinite communication cost. 
A strategy for making the amount of required communication finite and as small as possible 
is as follows. Instead of communicating directly the index $k$ [see Eq.~(\ref{map_psi_k})], 
Alice can communicate an amount of information that allows Bob to generate $k$ 
according to the probability distribution $\rho(k|y,\psi)$. By
Eq.~(\ref{cost_one-shot}), the minimal amount of required communication is
essentially equal to the capacity $C(K,Y|\Psi)$ of the channel $|\psi\rangle\rightarrow\{k,y\}$
(keep in mind that $y$ and $|\psi\rangle$ are uncorrelated).
If many simulations are performed in parallel, the reverse Shannon theorem implies that
there is a classical simulation such that the asymptotic communication cost
is strictly equal to $C(K,Y|\Psi)$.

In the following section, we will introduce a FC protocol working for measurements whose outcomes
are an arbitrary projector $|\phi\rangle\langle\phi|$ and its complement 
$\mathbb{1}-|\phi\rangle\langle\phi|$. We will denote by 
${\cal M}_2\equiv\{|\phi\rangle\langle\phi|,\mathbb{1}-|\phi\rangle\langle\phi|\}$
this kind of two-outcome measurement.

\section{Classical protocol for measurements ${\cal M}_2$}
\label{protocol}

In this section, we introduce an approximate completely $\psi$-epistemic model for
simulating a quantum channel with a restriction on the set of measurements $\cal M$.
More precisely, the model works for the two-outcome measurements 
${\cal M}_2=\{|\phi\rangle\langle\phi|,\mathbb{1}-|\phi\rangle\langle\phi|\}$, 
the outcomes being an arbitrary rank-$1$ projector $|\phi\rangle\langle\phi|$ and its 
complement. Using the procedure described in the previous
section, we will turn this protocol into a FC protocol, whose communication cost
does not depends on the number of communicated qubits and is inversely proportional
to the error.
The completely $\psi$-epistemic model is a higher-dimensional generalization of the 
exact Kochen-Specker (KS) model working for single qubits. Thus, we first review 
this latter model and, then, we introduce its generalization.

\subsection{Kochen-Specker model for single qubits}

The Kochen-Specker model can be seen as a classical protocol for simulating the 
communication of single qubits. This model does not use a shared 
random variable $y$. The communicated variable is a vector, say $|x\rangle$,
of the two-dimensional Hilbert space.
Given the quantum state $|\psi\rangle$, the sender, say Alice, prepares
$|x\rangle$ according to the probability distribution
\be\label{rho_KS}
\rho(x|\psi)=\frac{1}{2\pi^2}\left(|\langle x|\psi\rangle|^2-\frac{1}{2}\right)
\theta\left(|\langle x|\psi\rangle|^2-\frac{1}{2}\right),
\ee
where $\theta$ is the Heaviside step function. She sends $|x\rangle$
to a second party, say Bob.  The receiver then simulates
a projective measurement with two outcomes denoted by a pair of orthogonal
vectors, $|\phi\rangle$ and $|\phi_\perp\rangle$.
He generates the outcome $|\phi\rangle$ with probability
\be\label{cond_KS}
P(\phi|x)=\theta\left(|\langle x|\phi\rangle|^2-\frac{1}{2}\right)
\ee
This model simulates exactly a process of preparation, transmission and projective
measurement of a qubit, that is,
\be\label{equiv}
\int d^2x P(\phi|x)\rho(x|\psi)=|\langle\phi|\psi\rangle|^2,
\ee 
the right-hand side being the quantum probability of getting $|\phi\rangle$
given the quantum state $|\psi\rangle$.

In Ref.~\cite{montina_approx}, we used this model to derive a protocol that 
classically simulates the communication of a single qubit by using 2 bits of classical 
communication. This model differs from that reported in Ref.~\cite{toner}. Using the 
procedure described in the previous section, we also derived, from this model, a protocol 
working for simulations performed in parallel with asymptotic communication cost per
simulation equal to about $1.28$ bits~\cite{montina}, which is the capacity of the
channel $|\psi\rangle\rightarrow|x\rangle$ .

\subsection{Higher-dimensional generalization of KS model}

Now we present an approximate generalization of the KS model working in a
higher-dimensional Hilbert space and for measurements ${\cal M}_2$. Let us
denote by $N$ the Hilbert space dimension. The simplest generalization
of the KS model is as follows. The probability distribution $\rho(x|\psi)$,
previously defined by Eq.~(\ref{rho_KS}), becomes
\be\label{rho_gen}
\rho(x|\psi)=R(|\langle x|\psi\rangle|^2)
\theta\left(|\langle x|\psi\rangle|^2-\cos^2\theta_c\right),
\ee
where $R(\cdot)$ is a positive function such that probability distribution
$\rho(x|\psi)$ is normalized and $\theta_c$ is a parameter in the 
interval $[0,\pi/2]$. The communicated variable $|x\rangle$ is now 
a vector of the $N$-dimensional Hilbert space. The conditional probability
$P(\phi|x)$ defined by Eq.~(\ref{cond_KS}) takes now the more general form
\be\label{cond_gen}
P(\phi|x)=f(|\langle\phi|x\rangle|^2),
\ee
where $f(\cdot)$ is some function between zero and one. This model cannot
simulate exactly the quantum scenario, unless $\theta_c\rightarrow0$
($R$ going to infinite) and $f(y)=y$. In this limit case, the communicated
variable is the quantum state itself, that is, Alice sends the quantum
state to Bob, who uses it and the Born rule for evaluating the probability
of the outcome $|\phi\rangle$. This model cannot be turned into 
a FC protocol, as the capacity of the channel $|\psi\rangle\rightarrow|x\rangle$
is infinite. Thus, we keep $\theta_c$ different from zero and choose
$f(\cdot)$ and $R(\cdot)$ so that the error is as small as possible for 
any fixed $\theta_c$.
For a fixed $\theta_c$, we will see that the error scales as $N^{-1}$.

Because of the phenomenon of the concentration of the measure in high
dimension, there is a high probability that a vector $|x\rangle$ is
generated close to the contour of the support of $\rho(x|\psi)$,
that is, it is very likely that 
\be\label{contour}
|\langle x|\psi\rangle|^2\simeq\cos^2\theta_c.
\ee
Thus,
for high-dimensional Hilbert spaces, the function $R(\cdot)$ can
be approximated by a constant. Hereafter, for the sake of simplicity,
we assume that $R(\cdot)$ is actually a constant, $R_0$, determined by
the normalization of $\rho(x|\psi)$,
\be\label{R_const}
R(|\langle x|\psi\rangle|^2)=R_0.
\ee

An estimate of the function $f(\cdot)$ is given by the following reasoning.
By the concentration of the measure, it is possible to realize that, in 
high-dimensional Hilbert spaces, the vector $|x\rangle$ has also a high 
probability to be almost orthogonal to $|\phi\rangle-\langle\psi|\phi\rangle
|\psi\rangle$. This property and Eq.~(\ref{contour}) imply, that
\be
|\langle\phi|x\rangle|^2\simeq \cos^2\theta_c |\langle\phi|\psi\rangle|^2.
\ee
By this equation and Eqs.~(\ref{equiv},\ref{cond_gen}), we infer that
\be\label{estimate}
f(y)\simeq \frac{y}{\cos^2\theta_c},\text{  for  }
0<y<\cos^2\theta_c,
\ee
that is, the function $f(y)$ is very close to a linear function in the
interval $[0;\cos^2\theta_c]$.
This heuristic reasoning suggests a trial function $f$ of the form
$f(y)=c_0+c_1 y$, that is,
\be\label{trial_lin}
P(\phi|x)=c_0+c_1|\langle\phi|x\rangle|^2.
\ee

Initially, we assume that this linear form of $f(\cdot)$ holds in the whole domain 
$[0;1]$ of the function and will show that Eq.~(\ref{equiv}) is exactly satisfied
for a particular value of $c_0$ and $c_1$. Thus, the model reproduces exactly
quantum communication. However, 
this solution is not acceptable, since the conditional probability $P(\phi|x)$
turns out to be greater than $1$ or negative for some vectors $|\phi\rangle$.
This side effect is fixed by a slight change of the conditional probability
$P(\phi|x)$.
This correction introduces a small error, which will be evaluated.

The first step is to find the values of $c_0$ and $c_1$ such that Eq.~(\ref{equiv})
is exactly satisfied. From this equation and 
Eqs.~(\ref{rho_gen},\ref{R_const},\ref{trial_lin}), we have that

\be\label{exact_cond}
\begin{array}{c}
R_0\int d^{2N-2}x \left(c_0+c_1 |\langle x|\phi\rangle|^2\right) \hspace{1mm}
\theta\left(|\langle x|\psi\rangle|^2-\cos^2\theta_c\right) \vspace{1mm} \\
=|\langle\phi|\psi\rangle|^2.
\end{array}
\ee
Let us represent the vector $|x\rangle$ in the following coordinate
system,
\be
|x\rangle=\sin x_1 e^{i y_1}|\psi\rangle+\cos x_1\left[e^{i y_2}\sin x_2|1\rangle
+\cos x_2  e^{i y_3}|w\rangle\right],
\ee
where $|1\rangle$ is a vector orthogonal to $|\psi\rangle$ and lying in
the subspace spanned by $|\psi\rangle$ and $|\phi\rangle$, whereas
$|w\rangle$ is any vector orthogonal to $|\psi\rangle$ and $|1\rangle$.
The integration variables are $x_{1,2}$, $y_{1,2,3}$ and the $(N-2)$-dimensional
vector $|w\rangle$. The range of $x_i$ and $y_i$ is $[0;\pi/2]$ and 
$[0;2\pi]$, respectively. In this coordinate system, the measure of
an infinitesimal region is
\be
d^2x\,d^3y\, d^{2N-5}w \,\sin x_1 \cos^{2N-3}x_1\sin x_2\cos^{2N-5} x_2.
\ee
Using this measure and performing the integral in Eq.~(\ref{exact_cond}),
we find that,
\be
c_0+c_1\left[\cos^2\theta_c|\langle\phi|\psi\rangle|^2+\frac{1}{N}\sin^2\theta_c\right]=
|\langle\phi|\psi\rangle|^2.
\ee
This equation is satisfied for any $|\psi\rangle$ and $|\phi\rangle$ if
\be\begin{array}{c}
c_0=-\frac{1}{N}\tan^2\theta_c, \\ \vspace{1mm}
c_1=\cos^{-2}\theta_c.
\end{array}
\ee
Thus, from Eq.~(\ref{trial_lin}) we have that
\be\label{exact_sol}
P(\phi|x)=\frac{|\langle x|\phi\rangle|^2}{\cos^2\theta_c}-\frac{\tan^2\theta_c}{N},
\ee
which exhibits the correction term $N^{-1}\tan^2\theta_c$ with respect to
the heuristic estimate given by Eq.~(\ref{estimate}).
Although $P(\phi|x)$ satisfies exactly Eq.~(\ref{equiv}), 
however it is greater than one for $|\langle x|\phi\rangle|^2>\cos^2\theta_c
+N^{-1}\sin^2\theta_c$ and negative for 
$|\langle x|\phi\rangle|^2<N^{-1}\sin^2\theta_c$. Thus, this function
is not a conditional probability. The least invasive strategy for overcoming this
problem is to set the function equal to one or zero where it is greater than one
or negative, respectively. This makes the model approximate and we have to evaluate the 
introduced error. Summarizing, the approximate completely $\psi$-epistemic model
that generalizes the KS model is defined by the probability distributions
\be\label{flat_rho}
\rho(x|\psi)=R_0\hspace{0.5mm}\theta\left(|\langle x|\psi\rangle|^2-\cos^2\theta_c\right)
\ee
and
\be\label{approx_P}
P(\phi|x)=\left\{
\begin{array}{l}
1 \hspace{5mm}\text{ for } |\langle x|\phi\rangle|^2>\cos^2\theta_c+\frac{\sin^2\theta_c}{N} 
\vspace{2mm}  \\
0 \hspace{5mm}\text{ for } |\langle x|\phi\rangle|^2<\frac{\sin^2\theta_c}{N}
\vspace{2mm} \\
\frac{|\langle x|\phi\rangle|^2}{\cos^2\theta_c}-\frac{\tan^2\theta_c}{N} 
\hspace{4mm}\text{ elsewhere}
\end{array}
\right.
\ee
Hereafter we will consider the most relevant parameter region given by the inequality
\be\label{constr}
\tan^2\theta_c<N.
\ee
This condition simplifies the error analysis and rules out
only irrelevant protocols with error greater than $e^{-1}\simeq 0.36$.

Scrutinizing the exact quasi-probability distribution given by Eq.~(\ref{exact_sol}) 
and the approximate probability distribution, given by Eq.~(\ref{approx_P}), it is
easy to realize that the error of the protocol 
has two local maxima. One maximum, denoted by $\Delta_1$, is taken when 
$|\psi\rangle=|\phi\rangle$, the other one, say $\Delta_2$, when $|\psi\rangle$ and
$|\phi\rangle$ are orthogonal.

Let us evaluate $\Delta_1$. It is given by
\be
\Delta_1=1-\int dx P(\psi|x)\rho(x|\psi).
\ee
Performing the integral, we find under constraint~Eq.(\ref{constr}) that
\be
\Delta_1=\frac{1}{N}\left(1-\frac{1}{N}\right)^N\tan^2\theta_c\simeq
\frac{\tan^2\theta_c}{e N}.
\ee
The second local maximum is given by 
\be
\Delta_2=\int dx P(\psi_\perp|x)\rho(x|\psi),
\ee
where $|\psi_\perp\rangle$ is any vector orthogonal to $|\psi\rangle$.
For  $\tan^2\theta_c>(1-N^{-1})^{-1}$, we find that
\be
\Delta_2=\Delta_1-\frac{1}{N}\left(1-\frac{1}{N}-\cot^2\theta_c\right)^N\tan^2\theta_c,
\ee
otherwise $\Delta_2=\Delta_1$. Thus, $\Delta_1$ is always greater than or
equal to $\Delta_2$ and the absolute maximum error, say $\Delta$, is equal to
$\Delta_1$,
\be\label{error}
\Delta=\frac{1}{N}\left(1-\frac{1}{N}\right)^N\tan^2\theta_c\simeq \frac{1}{N}e^{-1}
tan^2\theta_c.
\ee

\subsection{$\psi$-epistemic FC protocol}
\label{completely_psi_epist}

The derived protocol is a completely $\psi$-epistemic model for $\theta_c\ne0$, that is,
the capacity of the channel $|\psi\rangle\rightarrow|x\rangle$ is finite. Let us
evaluate it. The mutual information $I(\Psi;X)$ is maximal for $\rho(\psi)$
constant. Since the
distribution $\rho(x|\psi)$ is uniform where it is different to zero, it is easy
to realize that the mutual information is the logarithm of the ratio between the volume 
of the space of unit vectors $|x\rangle$ and the volume of the support of 
$\rho(x|\psi)$, that is,
\be
I(X;\Psi)=\log_2\frac{\int dx\hspace{1mm}1}
{\int dx \hspace{1mm}\theta\left(|\langle x|\psi\rangle|^2-\cos^2\theta_c\right)}.
\ee
This equation gives
\be
I(X;\Psi)=-2(N-1)\log_2\left[\sin\theta_c\right].
\ee

According to the procedure described in section~\ref{procedure}, there is FC protocol 
whose asymptotic communication cost, ${\cal C}^{asym}$, is the mutual information, thus
${\cal C}^{asym}=-2(N-1)\log_2\left[\sin\theta_c\right]$. Using Eq.~(\ref{error}),
we can express the communication cost as a function of the error $\Delta$,
\be
{\cal C}^{asym}=(N-1)\log_2\left[1+\left(1-\frac{1}{N}\right)^N\frac{1}{N \Delta}\right].
\ee
Bearing in mind that the dimension $N$ grows exponentially with the number
of qubits, let us consider the relevant regime with $N\gg\Delta^{-1}$.
In this limit, we have that
\be\label{cost_psi_ep}
{\cal C}^{asym}\simeq\frac{1}{e\log_2 e}\Delta^{-1}\simeq \frac{0.255}{\Delta}.
\ee
Thus, the communication cost turns out to be independent of the number of
qubits and inversely proportional to the error in the high-dimensional
limit.

For single-shot simulations, the communication cost is bounded by Ineqs.~(\ref{cost_uniform}),
since the distributions $\rho(x|\psi)$ and $\rho(x)$ are uniform in their support.
Thus, the single-shot communication cost is equal to ${\cal C}^{asym}$ plus a possible
additional cost that is not greater than $\log_2 e\simeq 1.443$.

\subsubsection{Alternative protocol}
An alternative $\psi$-epistemic FC protocol can be derived using a dimensional
reduction strategy~\cite{brandao}. Since this strategy is quite known in quantum 
cryptography and the resulting protocol has a worse communication cost with respect
to the previous result, we just give a 
brief presentation of this protocol. The protocol is as follows. Alice and Bob share
a random unitary operator, say $\hat U$. Alice evaluates the normalized vector 
\be
|\psi_t\rangle\equiv \frac{\hat P\hat U|\psi\rangle}{\|\hat P\hat U|\psi\rangle\|},
\ee
where $\hat P$ is an operator projecting into a subspace with dimension $N_s$.
Similarly, Bob evaluates the vector
\be
|\phi_t\rangle\equiv \frac{\hat P\hat U|\phi\rangle}{\|\hat P\hat U|\phi\rangle\|}.
\ee
Alice approximates the vector $|\psi_t\rangle$ with a vector $|\psi_{net}\rangle$ of an 
$\epsilon$-net~\cite{enet}. 
An $\epsilon$-net is a set of vectors such that each vector of the Hilbert space
is within the distance $\epsilon$ of some vector in the set. Let us denote by $M$
the number of vector of the $\epsilon$-net.
There is an $\epsilon$-net such that
\be\label{M_eps}
M\propto\left(\frac{5}{\epsilon}\right)^{2N_s},
\ee
as proved in Ref.~\cite{hayden}. Alice then sends $|\psi_{net}\rangle$ to Bob. This
requires an amount of communication equal to
\be\label{cost_jl}
{\cal C}=\log_2 M.
\ee
Finally, Bob generates the outcome $|\phi\rangle\langle\phi|$ with probability
equal to $|\langle\psi_{net}|\phi_t\rangle|^2$. This protocol is an approximate
simulation of the quantum scenario. There are two sources of error. The first
one is the subspace projection. It introduces an error, say $\Delta_{proj}$ 
proportional to $N_s^{-1/2}$ and independent of the dimension $N$ of the
original Hilbert space~\cite{johnson}, 
\be\label{first_noise}
\Delta_{proj}\propto N_s^{-1/2}.
\ee
The second source is the
$\epsilon$-net, which introduces an error, $\Delta_{net}$, proportional to
$\epsilon$, so that Eq.~(\ref{M_eps}) can be written as
\be\label{M_delta}
M\propto\left(\frac{\alpha}{\Delta_{net}}\right)^{2N_s},
\ee
$\alpha$ being some constant.
From Eq.~(\ref{cost_jl},\ref{first_noise},\ref{M_delta}) we find that the communication cost, 
as a function of $\Delta_{net}$ and $\Delta_{proj}$, is
\be
{\cal C}\simeq \frac{\beta}{\Delta_{proj}^2}\log_2 \frac{\alpha}{\Delta_{net}},
\ee
where $\beta$ is a constant. This model is less efficient
than the previously derived model. Indeed, the communication cost grows as
$\Delta^{-2}\log_2\Delta^{-1}$, whereas the amount of communication in 
the previous model is proportional to $\Delta^{-1}$ [see Eq.~(\ref{cost_psi_ep})].

\subsection{Approximate $\psi$-ontic (strong) simulation}

In Sec.~\ref{completely_psi_epist}, we have presented a bounded-error $\psi$-epistemic
protocol whose communication cost is independent of the number of qubits and inversely
proportional to the worst-case error. 
Now, we compare this protocol with a bounded-error $\psi$-ontic protocol working for
the same quantum scenario and show
that the latter requires an amount of classical communication growing exponentially
with the number of qubits for a fixed worst-case error. 

In an exact $\psi$-ontic simulation, Alice sends the classical description of the
quantum state $|\psi\rangle$ to Bob. In other terms, Bob receives sufficient 
information to evaluate the probability $|\langle\phi|\psi\rangle|^2$ of any 
arbitrary event $|\phi\rangle\langle\phi|$. 
As defined in Sec.~\ref{class_sim}, in a bounded-error $\psi$-ontic protocol, 
Alice has to send an estimate, say $|\psi_e\rangle$, of the quantum state so that 
Bob can evaluate the probability of any event with an error bounded by a constant, 
say $\Delta$. This kind of simulation is also called strong simulation~\cite{nest}.
The vector $|\psi_e\rangle$ is chosen in an $\epsilon$-net of vectors
so that the distance between $|\psi_e\rangle$ and $|\psi\rangle$ cannot
be bigger than about $\Delta$. The number of elements of the $\epsilon$-net, say
$M$, scales exponentially with $N$. More precisely, 
\be
M\propto\left(\frac{\alpha}{\Delta}\right)^{2N}.
\ee
This scale law is optimal. Thus, the communication cost of the bounded-error
protocol is
\be
{\cal C}\simeq 2 N \log_2 \frac{\alpha}{\Delta}
\ee
It scales linearly with the Hilbert space dimension, that is, exponentially
with the number of qubits for a fixed worst-case error. Thus, there is
an exponential gap between the communication cost of the bounded-error 
$\psi$-epistemic protocol and the bounded-error $\psi$-ontic protocol.

\section{Conclusion}

There are two possible ways to simulate classically a
quantum channel. In the trivial way, the full classical description of the quantum
state is communicated by the sender to the receiver. In other words, the receiver
gets, in a single shot, the full information about the probabilities of every
event of any measurement.
This simulation, called $\psi$-ontic, requires
an infinite amount of communication. In the second way, the information about
the quantum state is encoded in the probability distribution of the communicated
variable. The receiver gets an amount of information that is sufficient 
to generate an event according to the quantum probabilities, but he does not get
the information about the quantum probabilities themselves. 
We have called this kind of protocol $\psi$-epistemic. In quantum computer
science, they are also known as weak simulations.

In this paper, we have presented a bounded-error $\psi$-epistemic protocol that classically 
simulates the communication of an arbitrary number $n$ of qubits with subsequent
measurement consisting of an arbitrary pure state projector and its complement.
The communication cost is independent of $n$ and inversely proportional to the 
worst-case error $\Delta$ in the limit $n\gg\Delta$. Conversely, a bounded-error 
$\psi$-ontic protocol requires an amount of classical communication growing 
exponentially with the number of qubits for a fixed error. Our model beats a previous
protocol based on the Johnson-Lindenstrauss lemma, whose communication cost
scales as $\Delta^{-2}\log_2\Delta^{-1}$, $\Delta$ being the error~\cite{brandao}.
The purpose of this work is to provide a further illustration that $\psi$-epistemic 
(weak) simulations of quantum systems can be more effective than $\psi$-ontic 
(strong) simulations. The still open challenge is to find exact completely $\psi$-epistemic 
theories of quantum systems or, equivalently, exact FC protocols, whose existence 
is still debated.
The state of the art about FC protocols is the lower bound $2^n-1$ for the
communication cost of a noiseless quantum channel with capacity $n$~\cite{monti_cap}.
In a following paper~\cite{mpw}, we will introduce a constructive procedure
to evaluate the communication complexity of general quantum channels. This procedure 
is based on the procedure used here, which was introduced in Ref.~\cite{montina}
and reviewed in Sec.~\ref{procedure}.

{\it Acknowledgments.} 
The author acknowledge useful discussions with Fernando Brandao.
Research at Perimeter Institute for Theoretical Physics is
supported in part by the Government of Canada through NSERC
and by the Province of Ontario through MRI.
This work is partially supported by the Swiss National 
Science Foundation, the NCCR QSIT, and the COST action on 
Fundamental Problems in Quantum Physics.

\bibliography{biblio.bib}

\end{document}